\documentclass[12pt]{article}
\usepackage{ascmac}
\usepackage{ulem}
\usepackage{latexsym}
\usepackage{amsmath}
\usepackage[top=3.5cm,bottom=3.5cm,left=2cm,right=2cm]{geometry}
\usepackage{pifont}          
\usepackage{amsmath,amssymb}
\usepackage[pdftex]{graphicx,color}
\usepackage{geometry}
\usepackage{caption}
\usepackage{ulem}
\usepackage{cancel}

\newcommand{\be}{\begin{equation}}
\newcommand{\e}{\end{equation}}
\newcommand{\aln}[1]{\begin{align}#1\end{align}}

\begin{document}

\title{
\vbox{
\baselineskip 14pt
\hfill \hbox{\normalsize KUNS-2763 
}} \vskip 1cm
\bf \Large  A note on higher spin symmetry in the IIB matrix model with the operator interpretation
\vskip 0.5cm
}
\author{
Katsuta~Sakai\thanks{E-mail: \tt katsutas@gauge.scphys.kyoto-u.ac.jp} 
\bigskip\\
\it \normalsize
 Department of Physics, Kyoto University, Kyoto 606-8502, Japan\\
\smallskip
}
\date{\empty}

\setcounter{page}{0}
\maketitle\thispagestyle{empty}
\abstract{\normalsize
We study the IIB matrix model in an interpretation where the matrices are differential operators defined on curved spacetimes. In this interpretation, coefficients of higher derivative operators formally appear to be massless higher spin fields. In this paper, we examine whether the unitary symmetry of the matrices includes appropriate higher spin gauge symmetries. We focus on fields that are bosonic and relatively simple in the viewpoint of the representation of Lorentz group. We find that the additional auxiliary fields need to be introduced in order to see the higher spin gauge symmetries explicitly. At the same time, we point out that a part of these extra fields are gauged-away, and the rest of part can be written in terms of a totally symmetric tensor field. The transformation to remove its longitudinal components exists as well. As a result, we observe that the independent physical DoF are the transverse components of that symmetric field, and that the theory describes the corresponding higher spin field. We also find that the field is not the Fronsdal field, rather the generalization of curvature. 
}
\newpage

\section{Introduction}\label{sec:intro}
To construct the theory of quantum gravity is one of the most important and difficult issues in the high-energy physics. When one discuss the issue within the framework of quantum field theory, the main obstacle is that there seems a strong tension between the unitarity and renormalizability in gravitational degrees of freedom (DoF). There are many attempts to prove the hidden healthiness of quantum gravity in field theory.\footnote{For example, asymptotic safety of gravity has been studied since \cite{Weinberg:1980gg}.}
On the other hand, the possibility that there is some UV completion of gravity other than ordinary field theory has been considered by numerous physicists. As is well known, string theory is one of the most promising candidate of the theory which contains quantum gravity. It describes the quantum graviton interacting other particles consistently. The shortage of the theory is that the spacetime is regarded as a coherent state of the graviton on some background,  and it does not completely explain the quantum dynamics of the spacetime. This problem is expected to be overcome by non-perturbative formulation of string theory.

The IIB matrix model \cite{Ishibashi:1996xs, Aoki:1998bq} is one of the candidates. It can be a constructive formulation of type IIB string theory. The matrix model is defined without a spacetime, and the DoF in it are several matrices. Nevertheless, it contains abundant physics involving string theory and spacetimes. The model correctly describes the force between parallel D-branes\cite{Ishibashi:1996xs}, and reproduces the light-cone Hamiltonian of string field theory\cite{Fukuma:1997en}. The spacetime emerges from the matrices. The emergent spacetime and the symmetry is discussed in \cite{Aoki:1998vn, Iso:1999xs}. It was reported in \cite{Kim:2011cr, Kim:2011ts} that a (3+1)-dimensional expanding universe appears from the path integral of the IIB matrix model, and the detailed feature about it has been studied\cite{Kim:2012mw, Ito:2015mem, Ito:2017rcr}. As for the fermionic sector, chiral zero-modes are induced with a particular backgrounds\cite{Nishimura:2013moa, Aoki:2014cya}. 
On the other hand, noncommutative spacetimes can emerge as well. They are solutions of the equations of motion in the model, and one can regard some parts of the fluctuations on them as the emergent gravity\cite{Steinacker:2007dq, Grosse:2008xr, Steinacker:2008ri}.\footnote{Some investigation of this scenario was made in \cite{Kawai:2016wfh} with the simplest background, i.e. noncommutative plane. Some other background provides more sophisticated dynamics\cite{Sperling:2018xrm, Sperling:2019xar}.}

Despite these variety of results, there is a room for discussion about the physical interpretation of the matrices. The most studies of the matrix model, like those mentioned above, treat the matrices as a ``coordinate DoF.'' It means that the expectation values or eigenvalues of the matrices are regarded as the coordinates in a flat spacetime, the existence of which is assumed. In this viewpoint, the distribution of such values forms some objects (strings, branes, universes, etc.) embedded in the flat spacetime. However, this interpretation is not the unique one with which one deals with the matrix model. Looking back on the form of the IIB matrix model, it is the large-$N$ reduction of super Yang-Mills theory. Roughly speaking, it implies that the DoF of the gauge field absorb their momenta. Then another natural option is to interpret the matrices as ``momenta DoF,'' and treat them as derivatives defined on some manifold. We call it {\it the operator interpretation}.

This interpretation was originally proposed in \cite{Hanada:2005vr, Kawai:2007zz}. One of its advantage is that one can obtain from the equations of motion curved spacetimes, on which the matrices as derivatives defined. Therefore it gives a background-independent formalism of emergent spacetimes in a self-consistent manner. Moreover, the $U(N)$ symmetry of the matrices are translated into a lot of symmetries of local fields, including diffeomorphism and local Lorentz symmetry. These facts suggest that the matrix model in the operator interpretation contains the DoF to describe the spacetime and gravity. On the other hand, one has to introduce many DoF, which, written in terms of local fields, formally appears to be massless higher spin fields. Although the $U(N)$ symmetry of the matrix model allows terms which are translated into mass terms for those fields, the quantum correction does not induces them at least one-loop level in the IIB matrix model\cite{Sakai:2017dxi}.

It is not clear whether these fields are actually physical DoF, and whether there are gauge symmetries which eliminate their potentially dangerous components, such as the longitudinal components of a vector field. In this paper, we investigate the symmetry of higher spin fields in some class, and see that the auxiliary fields need to be introduced in order to close the gauge transformation. There are gauge symmetries to remove the longitudinal components of the would-be spin-$s$ field and parts of the auxiliary fields. In addition, we pose some generalization of torsion-free conditions, which enable us to rewrite the rest parts of the auxiliary fields in terms of the physical field. As a result, we see that when we focus on the spin-$s$ fields, the gauge symmetries and the torsion-free conditions leave the transverse components of the fields in the totally symmetric representation.

The paper is organized as follows. In Section \ref{sec:review}, we have a review of the operator interpretation of the matrix model. Its properties and advantages are explained. In Section \ref{sec:higherspin} we study the higher spin symmetry in the IIB matrix model. This is our main analysis and observation. In Section \ref{sec:eom}, we discuss the symmetry at the level of dynamics. Finally in Section \ref{sec:summary}, we summarize the work and give some discussion.

\section{A review of the operator interpretation of the matrix model}\label{sec:review}
In this section, we quickly review the operator interpretation. For the detail see \cite{Hanada:2005vr, Kawai:2007zz}.

In the operator interpretation, one defines matrices as operators which act on some functional space. They are written as integral kernels. One can formally expand them to get an infinite series of differential operators. For example, when one considers operators acting on $C^\infty(\mathbb{R}^d)$, a matrix $K$ is represented as 
\aln{
K\in\mathrm{End}(C^\infty(\mathbb{R}^d)),&\nonumber\\
~~~~f(x)~\mapsto (K\cdot f)(x)& = \int_{\mathbb{R}^d}d^dyK(x,y)f(y)\nonumber\\
&=\int_{\mathbb{R}^d}d^dy\left[k(y)+k^\mu(y)\partial_\mu+k^{\mu\nu}(y)\partial_\mu\partial_\nu+\cdots\right]\delta(x-y)f(y)\nonumber\\
&=\left[k(x)+k^\mu(x)\partial_\mu+k^{\mu\nu}(x)\partial_\mu\partial_\nu+\cdots\right]f(x).\label{eq:diff_op}
}
In this sense, we interpret matrices as differential operators. When one discuss the choice of the functional space, one has to consider the concrete model and be careful in the physical picture.

In this paper we consider the IIB matrix model, the action of which is defined as
\aln{
S_\mathrm{IIB}=-\frac{1}{g^2}\mathrm{Tr}\left(\frac{1}{4}[A_a,A_b]^2+\frac{1}{2}\bar{\Psi}\Gamma^a[A_a,\Psi]\right). \label{eq:IIBmm}
}
Here, $A_a$ and $\Psi$ are $N\times N$ hermitian matrices with Lorentz and spinor indices, respectively.  In this model, we attempt to treat them as differential operators with the indices, which act on the functional space defined on some curved spacetime manifold $\mathcal{M}$. Naively, it appears to be realized by representing $A_a$ and $\Psi$ as Eq.(\ref{eq:diff_op}), with derivatives replaced by the covariant derivatives and the indices reinterpreted as those for the local Lorentz transformation: 
\aln{
A_a:f(x)\mapsto(A_a\cdot f)(x)= \left[a_a(x)+\frac{1}{2}[a_a^{~\mu}(x)\,i\nabla_\mu]_h+\frac{1}{2}[a_a^{~\mu\nu}(x)\,i\nabla_\mu\,i\nabla_\nu]_h+\cdots\right]f(x),\label{eq:diff_op_A}
}
where $f(x)\in C^\infty(\mathcal{M})$, and $[~~~]_h$ is an order-symmetrized product introduced to guarantee the hermicity of the matrix; $[X_1X_2\cdots X_n]:=X_1X_2\cdots X_n+X_n\cdots X_2X_1$.

However, the above treatment actually faces some obstacles. There are two problems involving the explicit indices of the matrices. First, the operation of the matrix adds to a function the Lorentz index, and changes their representation of Lorentz group. Thus the operation of a matrix is not closed on the functional space of the specific representation. Even when one considers the space of functions of all representation, one cannot consider invertible operators since acting a matrix is always increase the rank of representation. 
It means that we cannot interpret matrices as endmorphisms on the space, conflicting to the fact that the matrices themselves are so.

Another problem is that the operator in the right hand side of Eq.(\ref{eq:diff_op_A}) should be a vectorial operator. On a curved manifold, components of a vector is first defined in each local coordinate patch, and is then glued with a transition function in the overlap regions, to form a vector consistently. In contrast, each matrix such as $A_1,A_2$ or $A_3$, is defined independently. This fact indicates that $A_a$ should not be regarded as a vector.

In order to overcome these difficulty, we introduce the principal $SO(d-1,1)$ bundle $E_\mathrm{prin}$ the base space of which is the curved spacetime $\mathcal{M}$.\footnote{To be exact, one have to introduce $Spin(d)$ bundle. In this paper, the difference is not important since we discuss the local property of the group only.} Locally, the bundle is written as the direct product of $\mathcal{M}$ and the Lorentz group $G=SO(d-1,1)$. Then functions on it depend on the coordinates of spacetime $x$ and those of Lorentz group $g$. They are in the regular representation of the Lorentz group: 
\aln{
f(x,g)&\in E_{\mathrm{prin}},\\
G\ni h:f(x,g)&\mapsto (h\cdot f)(x,g)=f(x,h^{-1}g)
}
The important property of the regular representation is that the tensor product of it and arbitrary irreducible representation is isomorphic to the direct sum of the regular representations. When we denote the vector space for the regular representation and an irreducible representation as $V_{\mathrm{reg}}$ and $V_\mathrm{r}$, respectively, the property is written as 
\aln{
V_{\mathrm{r}}\otimes V_\mathrm{reg}\simeq V_\mathrm{reg}\oplus\cdots\oplus V_\mathrm{reg}.
}
Here the number of $V_\mathrm{reg}$ in the right hand side is the same as the dimension of $V_\mathrm{r}$. In terms of the functions on $E_\mathrm{prin}$, this isomorphism is realized as below. When we denote the function in the product representation as $f_i(x,g)$, with $i$ being the index for r-representation, its transformation law is written as 
\aln{
G\ni h:f_i(x,g)\mapsto R_i^{\left<r\right>j}(h)f_j(x,h^{-1}g).
}
Here, $R_i^{\left<\text{r}\right>j}(h)$ is the Lorentz group matrix in r-representation. Then, the isomorphism is realized by separating the matrix from the function: 
\aln{
f_{i}(x,g)=:R_i^{\left<\text{r}\right>(j)}(g)f'_{(j)}(x,g)
} 
Here, $f'_{(i)}$ belongs to the regular representation, and $R_i^{\left<\text{r}\right>(j)}(g)$ plays a roll of Crebsh-Gordan coefficients. The components for index $(i)$ do not mix by the Lorentz transformations. Indeed, since $f'_{(i)}(x,g)=R^{\left<r\right>j}_{(i)}(g^{-1})f_{j}(x,g)$, its transformation law is given by
\aln{
f'_{(i)}(x,g)&\mapsto(R^{\left<\text{r}\right>}(g)^{-1})_{(i)}^{~~j}(h\cdot f)_{j}(x,g)\nonumber\\
&=R_{(i)}^{\left<\text{r}\right>\,j}(g^{-1}) R_j^{\left<\text{r}\right>k}(h)f_k(x,h^{-1}g)\nonumber\\
&=R^{\left<\text{r}\right>\,j}_{(i)}((h^{-1}g)^{-1})f_{j}(x,h^{-1}g)\nonumber\\
&=f'_{(i)}(x,h^{-1}g).
}
In the following, we represent indices that are not affected by the transformation with parentheses, such as $(i)$.

With the above statement for the regular representation, it is easy to see that one can interpret matrices as differential operators acting on $C^\infty(E_\mathrm{prin})$, with indices in parentheses. We rewrite $A_a$ in the matrix model Eq.(\ref{eq:IIBmm}) as $\hat{A}_{(a)}$, 
\aln{
S_\mathrm{IIB}=-\frac{1}{g^2}\mathrm{Tr}\left(\frac{1}{4}[\hat{A}_{(a)},\hat{A}_{(b)}]^2+\frac{1}{2}\bar{\Psi}\Gamma^a[\hat{A}_{(a)},\Psi]\right), \label{eq:IIBmm2}
}
and the matrices take the form of
\aln{
\hat{A}_{(a)}&=R^{\left<\text{v}\right>\,b}_{(a)}(g^{-1})A_b(x,g),\\
 A_b(x,g)&=a_b(x,g)+\frac{1}{2}[a_b^{~\mu}(x,g)\,i\nabla_\mu]_h+\frac{1}{2}[a_a^{~\mu\nu}(x,g)\,i\nabla_\mu\,i\nabla_\nu]_h+\cdots, \label{eq:diff_op_A2}
}
With $R^{\left<\text{v}\right>\,b}_{(a)}$ is the matrix of the vector representation. This operator does not change the representation of functions (from the regular representation to itself). Therefore $\hat{A}_{(a)}\in \mathrm{End}(C^\infty(E_\mathrm{prin}))$. Furthermore, each component of $\hat{A}_{(a)}$ is a scalar operator, in the sense that they do not mix under the Lorentz transformation, and hence under the operation of the transition function. Thus each of the operators $\hat{A}_{(1)}, \hat{A}_{(2)},\cdots$ is independently defined, as each matrices should be.

Once we adopt such a interpretation, it is natural that we extend the class of operators to that of the general derivative operators defined on $E_\mathrm{prin}$. It means that we deal with operators involving the derivatives with respect both to the spacetime and Lorentz group coordinates. The latter is equivalent to the Lorentz generators for the regular representation $\mathcal{O}_{cd}$. Then we identify $A_b(x,g)$ in Eq.(\ref{eq:diff_op_A2}) to the operators of the following form:
\aln{
A_b(x,g)=&a_b(x,g)+\frac{1}{2}[a_b^{~\mu}(x,g)\,i\nabla_\mu]_h+\frac{1}{2}[a_a^{~\mu\nu}(x,g)\,i\nabla_\mu\,i\nabla_\nu]_h+\cdots\nonumber\\
&+\frac{1}{2}[a_b^{~cd}(x,g)\,\mathcal{O}_{cd}]_h+\frac{1}{2}[a_b^{~\mu cd}(x,g)\,\mathcal{O}_{cd}\,i\nabla_\mu]_h+\cdots\nonumber\\
&+\frac{1}{2}[a_b^{~cc'dd'}(x,g)\,\mathcal{O}_{cd}\,\mathcal{O}_{c'd'}]_h+\cdots. \label{eq:diff_op_A3}
} 
Moreover, it is notable that each field such as $a_b(x,g)$ and $a_b^{~\mu}(x,g)$ is decomposed to the infinite series of the field in the representation that is the product of irreducible ones and their conjugates: 
\aln{
a_b(x,g)=\sum_{\text{r:irreducible}}R^{\left<\text{r}\right>\,j}_{i}(g)a_{bj}^{~~i}(x)
}
As a result, we obtain a numerous infinite number of fields in various representation.

The advantage of the operator interpretation is that we can describe curved spaces in a background-independent manner. In the following, we focus on the bosonic part of the IIB matrix model Eq.(\ref{eq:IIBmm2}), {\it i.e.} we restrict the analysis with the condition $\Psi=0$. When we take an ansatz $A_a(x,g)=i\nabla_a$, the equations of motion is rewritten as 
\aln{
&[\hat{A}^{(b)},[\hat{A}_{(b)},\hat{A}^{(a)}]]=0\nonumber\\
\Leftrightarrow~&[A^b,[A_b,A^a]]=0\nonumber\\
\Leftrightarrow~&[\nabla^b,[\nabla_b,\nabla^a]]=0\nonumber\\
\Leftrightarrow~&\nabla^cR^{da}\mathcal{O}_{cd}-R^{a\mu}\nabla_\mu=0\nonumber\\
\Leftrightarrow~&R^{ab}=0.
\label{eq:eomoriginal}
}

Here $R^{ab}$ is the Ricci tensor, and thus we have obtained the vacuum solutions of Einstein equation. This solution settles the base spacetime of $E_\text{prin}$ self-consistently. If we add to the ansatz a field in the vector representation, $A_a(x,g)=a_a(x)+i\nabla_a$, we can show that the EOM for $a_a(x)$ is Maxwell equation on the base spacetime.

Another important advantage of the interpretation is that the model possesses the manifest symmetries including those of the diffeomorphism and local Lorentz. Eq.(\ref{eq:IIBmm2}) has $U(N)$ symmetry, $\delta \hat{A}^{(a)}=i[\hat{A}^{(a)},\Lambda]$ with $\Lambda$ being an hermitian matrix. Rewriting it in terms of operators and choosing the specific form of $\Lambda$, we obtain such symmetries. For example, let us treat the spacetime DoF with the parametrization
\aln{
A_a(x,g)=i\nabla_a=i e_a^{~\mu}(\partial_\mu+i\omega_\mu^{~cd}\mathcal{O}_{cd}).  
}
When we choose the transformation parameter as $\lambda=(1/2)[\lambda^\mu(x)\,i\partial_\mu]_h$, then the transformation laws for each field are given by 
\aln{
\delta e_a^{~\mu} &=-\lambda^\nu\partial_\nu e_a^{~\mu}+e_a^{~\nu}\partial_\nu\lambda^\mu,\\
\delta \omega_\mu^{~cd}&=-\lambda^\nu\partial_\nu\omega_\mu^{~cd}. 
}
On the other hand, another choice of the parameter $\lambda=\lambda^{c'd'}(x)\mathcal{O}_{c'd'}$ yields the transformation laws below: 
\aln{
\delta e_a^{~\mu}&=-\lambda_a^{~b}e_b^{~\mu},\\
\delta \omega_\mu^{~cd}&=\partial_\mu\lambda^{cd}+2\lambda^{[c}_{~~e}\omega_\mu^{~d]e}.
}
The square bracket represents anti-symmetrization of the indices. These transformation laws allow us to identify $e_a^{~\mu}$ and $\omega_\mu^{~cd}$ to the vielbein and spin connection, respectively. At the same time, the transformations are diffeomorphism and local Lorentz transformation, respectively. When one consider the vector field $a_a(x)$, it transforms as an $U(1)$ gauge field $\delta a_a = -\partial_a\lambda$ with the gauge parameter $\Lambda=\lambda(x)$.

The two advantages suggest that we can describe the spacetime and gravity DoF by the IIB matrix model in the operator interpretation. In this interpretation, the effective action takes the form of a polynomial of what we usually treat as an action\cite{Asano:2012mn}. That unusual effective action is supposed to yield the solution of fine-tuning problem\cite{Kawai:2013wwa, Hamada:2015dja, Kawana:2016tkw}. Note that the matrix model contains infinitely many fields. They are generically of higher-rank representation, and are all massless at the tree level or in the presence of the supersymmetry\cite{Sakai:2017dxi}. Therefore, we expect them to be massless higher-spin fields with appropriate gauge symmetry.

\section{Higher spin gauge symmetries in the IIB matrix model}\label{sec:higherspin}
Although the IIB matrix model is likely to contain higher spin fields, it remains to be justified that they describe physically healthy DoF. More concretely, it is unclear so far whether there are abundant gauge DoF in $U(N)$ transformation of matrices to eliminate the longitudinal components of the fields. In this section, we investigate such aspect of the matrix model. In the following, we focus on a restricted class of the fields, namely the bosonic fields that are independent of group coordinates $g$. Thus they are not in the product representation of the tensor one and the regular one.

In our analysis, we treat operators in the ``semi-classical limit.'' Namely, we replace the derivatives with some c-numbers and define a Poisson bracket corresponding the commutator:
\aln{
&~~~~~~~~~~~~~~~~\partial_\mu\rightarrow p_\mu,~~\mathcal{O}_{ab}\rightarrow t_{ab},\\
&~~~~~~~~~~~~~~~~~~~~~~i[\cdot,\cdot]\rightarrow \{\cdot,\cdot\},\nonumber\\
&\{p_\mu, x^\nu\}=\delta_\mu^\nu,\nonumber\\
&\{t_{ab},g_{ij}\}=i(M_{ab}g)_{ij},~~\{t_{ab},t_{a'b'}\}=if_{ab,a'b',cd}t_{cd}.
}
Here $M_{ab}$ denotes the Lorentz generator in the fundamental representation, and $f_{ab,a'b',cd}t_{cd}$ is the structure constant. This limit enables us to ignore the order of the derivatives and coordinates in the expansion of $A_a(x,g)$, and simplifies the analysis.

In the ordinary field theory, a massless spin-$s$ field is described by a rank-$s$ symmetric double-traceless tensor field\cite{Fronsdal:1978}: 
\aln{
a_{\mu(s)}(x)~~~\mathrm{s.t.} ~~~a^{\nu_1\nu_2}_{~~~~~\nu_1\nu_2\mu(s-4)}=0, 
\label{eq:ordinaryhigherspin}
}
where $\mu(s)$ denotes the symmetrized indices $(\mu_1\cdots\mu_{s})$.\footnote{In the spin-$3$ case, any tracelessness is not imposed.} The gauge transformation of it is written as 
\aln{
\delta a_{\mu(s)}=\partial_\mu\lambda_{\mu(s-1)}
}
with $\lambda_{\mu(s-1)}$ is a rank-$(s-1)$ symmetric traceless tensor parameter. We formally express the symmetrized indices the same letter.

\begin{figure}
\begin{center}
\includegraphics[width=11cm]{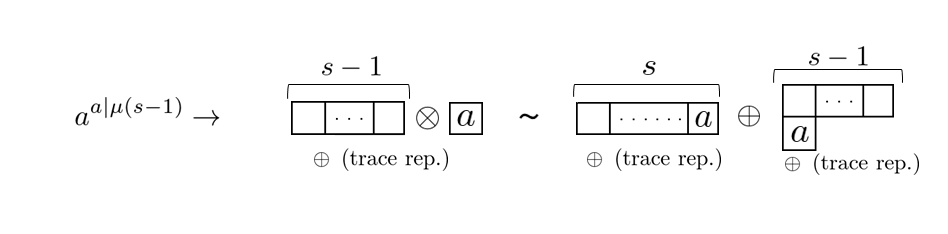}
\caption{The representational structure of the field. The bar between the indices in the field denotes the tensor product, as mentioned below. The field is of a tensor product of vector representation and rank-$(s-1)$ symmetric one. It is decomposed into rank-$s$ symmetric representation and the rest ``hook-type'' one. Note that all the representations
contain the trace part, and they are reducible.}
\label{fig:Youngtableaux1}
\end{center}
\end{figure}

Naively, it seems natural that a spin-$s$ field in the flat spacetime background is described in the operator interpretation as 
\aln{
A^a=p^a + a^{a\mu(s-1)}(x)p_{\mu(s-1)},\label{eq:spin-s}
}
where $p_{\mu(s-1)}:=p_{(\mu_1}\cdots p_{\mu_{s-1})}$. The first term in Eq.(\ref{eq:spin-s}) is for the background. We attempt to find the appropriate gauge transformation for the field. First, the most simple gauge parameter we have is the following form: 
\aln{
\Lambda = \lambda^{\mu(s-1)}(x)p_{\mu(s-1)},
}
which realizes the transformation
\aln{
\delta \hat{A}^{(a)}&=\{\hat{A}^{(a)},\Lambda\}\nonumber\\
\Leftrightarrow~~\delta a^{a\mu(s-1)}&=\partial^a\lambda^{\mu(s-1)}+O(a\times\lambda).
\label{eq:gaugetrsf1}
}
In the analysis, we will ignore the second term in the right hand side of Eq.(\ref{eq:gaugetrsf1}). Although the validity of it needs to be analyzed, in this paper we assume that the discussion around the elimination of the DoF can be held focusing only on the inhomogeneous term. Of course, Eq.(\ref{eq:gaugetrsf1}) is not sufficient for the elimination of the longitudinal components of $a^{a\mu(s-1)}$, because it includes non-totally symmetric tensor components. It comes from the fact that $a^{a\mu(s-1)}$ behave as the product representation of the vector one (having the index $a$) and rank-$(s-1)$ symmetric tensor one ($\mu(s-1)$). The representation is decomposed into two representations and their traces (Fig.(\ref{fig:Youngtableaux1})). The extra components are the two-row representation tensor, characterized by the second tableaux in Fig.(\ref{fig:Youngtableaux1}). 
%

In terms of the field, we rewrite $a^{a\mu(s-1)}$ as $a^{a|\mu(s-1)}$ and the decomposition as 
\aln{
a^{a|\mu(s-1)}=h^{a\mu(s-1)}+b^{a,\mu(s-1)}.
}
From now on, we separate the indices for tensor products by bars, and for different rows in the Young tableaux by commas. The sequence of indices without commas or bars are symmetric.

The problem is whether there is any gauge transformation to remove $b^{a,\mu(s-1)}$. We take a new gauge parameter in the following form: 
\aln{
\Lambda=\lambda^{\mu(s-1)}p_{\mu(s-1)}+\lambda^{c,d\mu(s-2)}t_{cd}\,p_{\mu(s-2)}. 
}
With this parameter we get the transformation law as below: 
\aln{
\delta A^a = \partial^a\lambda^{\mu(s-1)}p_{\mu(s-1)}-\frac{1}{2}(\lambda^{a,\mu(s-1)}-\lambda^{\mu,a\mu(s-2)})p_{\mu(s-1)}+\partial^a\lambda^{c,d\mu(s-2)}t_{cd}\,p_{\mu(s-2)}.
\label{eq:opgaugetrsf2}
}
In terms of the fields this is written as
\aln{
\delta h^{a\mu(s-1)}&=\partial^{(a}\lambda^{\mu(s-1))},\\
\delta b^{a,\mu(s-1)}&=-\frac{s}{2(s-1)}\lambda^{a,\mu(s-1)}+(\partial^a\lambda^{\mu(s-1)})_{\mathbb{P}(s-1,1)}.
\label{eq:gaugetrsf2}
}
In the above equations, $\mathbb{P}(m,n)$ represents the projection into the representation for the Young tableaux which consists of  an $m$-boxes row and an $n$-boxes row. The coefficient $s/(s-1)$ appears from the normalization of the projection, $(\lambda^{a,\mu(s-1)})_{\mathbb{P}(s-1,1)}=\lambda^{a,\mu(s-1)}$. Eq.(\ref{eq:gaugetrsf2}) indicates that we can remove all the components of $b^{a,\mu(s-1)}$ by this transformation. Furthermore, we can remove the longitudinal components of totally symmetric tensor $a^{a\mu(s-1)}$ with $b^{a,\mu(s-1)}$ kept zero by choosing $\lambda^{a,\mu(s-1)}$ appropriately.

However, Eq.(\ref{eq:opgaugetrsf2}) includes the extra change of $A^a$, {\it i.e.} the third term on the right hand side. In order to close the transformation law, it is necessary to introduce new DoF. Therefore we are forced to consider the operator of the following form:
\aln{
A^a=p_a + a^{a|\mu(s-1)}(x)p_{\mu(s-1)}+\omega^{a|c,d\mu(s-2)}t_{cd}\,p_{\mu(s-2)},\label{eq:spin-s2}
}
where $\omega^{a|c,d\mu(s-2)}$ is an additional field. Then we have again the problem of whether $\omega^{a|c,d\mu(s-2)}$ can be removed by any gauge transformation.

Before discussing the gauge transformation, note that $\omega^{a|c,\mu(s-1)}$ is seen as a higher spin counterpart for the spin connection. In the spin-2 case, the spin connection $\omega^{a|b,c}$ is written in terms of vielbein through the torsion-free condition
\aln{
T^{a|bc}=\partial^be^{c|a}-\partial^ce^{b|a}+\omega^{c|d,a}e_{d|}^{~~b}-\omega^{b|d,a}e_{d|}^{~~c}=0.
\label{eq:torsionfreecond}
}  
Keeping this fact in mind, we shall pose the generalized torsion-free condition: 
\aln{
\frac{2(s-1)}{s}(\partial^ba^{c|\mu(s-1)}-\partial^ca^{b|\mu(s-1)})+\omega^{b|c,\mu(s-1)}-\omega^{c|b,\mu(s-1)}=0.
\label{eq:gtorsionfreecond}
}
In spin 2 case, this coincides with Eq.(\ref{eq:torsionfreecond}) with the vielbein being small fluctuation around the flat space. 
The general solution of Eq.(\ref{eq:gtorsionfreecond}) is written as 
\aln{
\omega^{a|b,\mu(s-1)}=\frac{s-1}{s}&\Bigl(\partial^ba^{a|\mu(s-1)}-\partial^aa^{b|\mu(s-1)}+\partial^ba^{\mu|a\mu(s-2)}-\partial^\mu a^{b|a\mu(s-2)}\nonumber\\
&~~~~~+\partial^aa^{\mu|b\mu(s-2)}-\partial^\mu a^{a|b\mu(s-2)}\Bigr)+\zeta^{ab,\mu(s-1)},
}
where $\zeta^{ab,\mu(s-1)}$ is an arbitrary tensor corresponding to the Young tableaux  whose two rows consist of $(s-1)$ and 2 boxes, respectively. 
Therefore the additional field $\omega^{a|b,\mu(s-1)}$ is written with $a^{a|\mu(s-1)}$ through the above equation, except for components of $\zeta^{ab,\mu(s-1)}$.

Fortunately, it is possible to eliminate $\zeta^{ab,\mu(s-1)}$ by another gauge transformation. we choose a gauge parameter of the form below: 
\aln{
\Lambda=\lambda^{\mu(s-1)}p_{\mu(s-1)}+\lambda^{c,d\mu(s-2)}t_{cd}\,p_{\mu(s-2)}+\lambda^{c(2),d(2)\mu(s-3)}t^2_{cd}\,p_{\mu(s-3)}, 
}
with the notations are defined as
\aln{
t^n_{cd}:=t_{c_1d_1}\cdots t_{c_nd_n}.
}
The gauge transformation of $A^a$ is then
\aln{
\delta A^a = &\partial^a\lambda^{\mu(s-1)}p_{\mu(s-1)}-\frac{1}{2}(\lambda^{a,\mu(s-1)}-\lambda^{\mu,a\mu(s-2)})p_{\mu(s-1)}\nonumber\\
&+\partial^a\lambda^{c,d\mu(s-1)}t_{cd}\,p_{\mu(s-2)}-\frac{1}{2}(\lambda^{ac,d\mu(s-2)}-\lambda^{\mu c,ad\mu(s-3)}+\lambda^{ca,d\mu(s-2)}-\lambda^{c\mu,da\mu(s-3)})t_{cd}\,p_{\mu(s-2)}\nonumber\\
&+\partial^a\lambda^{c(2),d(2)\mu(s-3)}t^2_{cd}\,p_{\mu(s-3)},
\label{eq:opgaugetrsf3}
}
hence
\aln{
\delta a^{a|\mu(s-1)}&=\partial^a\lambda^{\mu(s-1)}-\frac{s}{2(s-1)}\lambda^{a,\mu(s-1)},
\label{eq:gaugetrsf3a}\\
\delta \omega^{a|c,d\mu(s-2)}&=\partial^a\lambda^{c,d\mu(s-2)}-\frac{s}{4(s-2)}\lambda^{ac,d\mu(s-2)}.
\label{eq:gaugetrsf3b}
}
Eq.(\ref{eq:gaugetrsf3a}) is equivalent to Eqs.(\ref{eq:gaugetrsf2}), while Eq.(\ref{eq:gaugetrsf3b}) is consistent with the imposed condition Eq.(\ref{eq:gtorsionfreecond}). As a result, a part of $\omega^{a|c,d\mu(s-2)}$ can be removed by the second term in Eq.(\ref{eq:gaugetrsf3b}), and the rest part is written in terms of $a^{a|\mu(s-1)}$. Therefore, there is no independent DoF in $\omega^{a|c,d\mu(s-2)}$.

Due to the last term in Eq.(\ref{eq:opgaugetrsf3}), we have to introduce further additional field in order to close the gauge transformation. Remarkably, the present discussion is somewhat similar to that of the higher spin gauge theory in form language\cite{Vasiliev:1980as}.\footnote{For a review see \cite{Didenko:2014dwa}.} In the viewpoint of the gauge transformation, we find that the present analysis can be done almost in parallel with the study in \cite{Engquist:2007yk}, although the generalized torsion-free conditions are different. Therefore, we state the discussion briefly. In order to close gauge transformation completely, we have to consider the operator of the following form:
\aln{
A^a=p^a+a^{a|\mu(s-1)}p_{\mu(s-1)}+\sum_{n=1}^{s-1}\omega^{a|c(n),d(n)\mu(s-1-n)}t_{cd}^{\,n}\,p_{\mu(s-1-n)}
}
Appropriate gauge parameter is given by
\aln{
\Lambda=\lambda^{\mu(s-1)}p_{\mu(s-1)}+\sum_{n=1}^{s-1}\lambda^{c(n),d(n)\mu(s-1-n)}\,t_{cd}^{\,n}\,p_{\mu(s-1-n)},
}
which leads to the transformation laws
\aln{
\delta a^{a|\mu(s-1)}&=\partial^a\lambda^{\mu(s-1)}-\frac{s}{2(s-1)}\lambda^{a,\mu(s-1)},
\label{eq:gaugetrsf4a}\\
\delta \omega^{a|c(n),d(n)\mu(s-1-n)}&=\partial^a\lambda^{c(n),d(n)\mu(s-1-n)}-\frac{s}{2^n(s-1-n)}\lambda^{ac(n),d(n)\mu(s-1-n)}.~~(1\leq n\leq s-2),
\label{eq:gaugetrsf4b}\\
\delta\omega^{a|c(s-1),d(s-1)}&=\partial^a\lambda^{c(n),d(n)}.
\label{eq:gaugetrsf4c}
}
Now we impose a set of generalized torsion-free conditions
\aln{
&\frac{2^n(s-1-n)}{s}(\partial^a\omega^{b|c(n),d(n)\mu(s-1-n)}-\partial^b\omega^{a|c(n),d(n)\mu(s-1-n)})\nonumber\\
&~~~~~~~~~~~~~~~~~~+\omega^{a|bc(n),d(n)\mu(s-1-n)}-\omega^{b|ac(n),d(n)\mu(s-1-n)}=0.~~(1\leq n\leq s-2)
\label{eq:gtorsionfreecondfull}
}
Due to this equations, a part of each extra fields $\omega^{a|c(n),d(n)\mu(s-1-n)}$ is written in terms of the ``lower'' extra fields recursively. At the same time, the rest part of $\omega^{a|c(n),d(n)\mu(s-1-n)}$ can be removed with the gauge transformation, in particular with the second term in Eq.(\ref{eq:gaugetrsf4b}). As for the highest extra field $\omega^{a|c(s-1),d(s-1)}$, there is no gauge parameter with which we can eliminate the DoF of the field. However, the generalized torsion-free condition for it can be solved and the whole part of it is expressed with $\omega^{a|c(s-2),d(s-2)\mu}$ without ambiguity: 
\aln{
\omega^{a|bc(s-2),d(s-1)}=-\frac{1}{2}\biggl[&
\partial^a\omega^{b|c(s-2),d(s-1)}-\partial^b\omega^{a|c(s-2),d(s-1)}\nonumber\\
&-(s-1)\Bigl(
\partial^d\omega^{a|c(s-2),bd(s-2)}-\partial^a\omega^{d|c(s-2),bd(s-2)}\nonumber\\
&~~~~~~~~~~~~~~-\partial^b\omega^{d|c(s-2),ad(s-2)}-\partial^d\omega^{b|c(s-2),ad(s-2)}\nonumber\\
&~~~~~~~~~~~~~~+\partial^d\omega^{a|bc(s-3),cd(s-2)}-\partial^a\omega^{d|bc(s-3),cd(s-2)}
\Bigr)
\biggr]
}
In the derivation of the above equation, we have made use of the Bianchi identity
\aln{
\omega^{a|c(s-1),cd(s-2)}=\omega^{a|dc(s-2),d(s-1)}=0, 
}
and a relation which is derived from it, 
\aln{
\omega^{a|dc(s-2),bd(s-2)}=-\frac{1}{s-1}\omega^{a|bc(s-2),d(s-1)}. 
}
The fact that the $\omega^{a|c(s-1),d(s-1)}$ can be solved is on the same foot as that the spin connection can be solved in terms of the vielbein.

According to these discussion, we can conclude that $b^{a,\mu(s-1)}$ and all the extra fields $\omega^{a|c(n),d(n)\mu(s-1-n)}$ are eliminated either with gauge transformation or with generalized torsion-free condition. In this sense, the extra fields are auxiliary fields. Furthermore, we can still remove the longitudinal component of $h^{a\mu(s-1)}$ by an appropriate gauge transformation. It is driven both by the parameter $\lambda^{\mu(s-1)}$ and the higher rank parameters $\lambda^{c(n),d(n)\mu(s-1-n)}$. The former removes the longitudinal components directly, while the latter compensate the change in $\omega^{a|c(n),d(n)\mu(s-1-n)}$ and keep them zero. Therefore, we are left the transverse component of $h^{a\mu(s-1)}$ as the only physical DoF.

After gauge-fixing and eliminating fields except $h^{a\mu(s-1)}$, the matrices take the following form:
\aln{
A^a=p^a+\sum_{n=0}^{s-1}\frac{1}{n!}\partial^{c(n)}h^{d(n)a\mu(s-1-n)}t_{cd}^n~p_{\mu(s-1-n)}.
\label{eq:gaugefixedA}
}
On the other hand, the explicit form of the residual gauge degrees of freedom which remove the longitudinal components of $h^{a\mu(s-1)}$ is written as
\aln{
\Lambda=\sum_{n=0}^{s-1}\frac{s-n}{n!}\partial^{c(n)}\lambda^{d(n)\mu(s-1-n)}t_{cd}^n~p_{\mu(s-1-n)}.
}
Then we find that the unitary transformation of matrices is equivalent of a higher spin gauge transformation:
\aln{
\delta A^a=\{A^a,\Lambda\}~~\Leftrightarrow~~\delta h^{a\mu(s-1)}=\partial^{(a}\lambda^{\mu(s-1))}.
}

Here $h^{a\mu(s-1)}$ does not belong to an irreducible representation, since it contains the trace part. In this respect, there is some difference between the field and the ordinary higher spin fields, which satisfies the double-traceless condition Eq.(\ref{eq:ordinaryhigherspin}). In the ordinary case, the condition is required to make the theory gauge-invariant, with the gauge parameter being traceless. As for our case, we already have gauge invariance with the traceful field $h^{a\mu(s-1)}$ and the parameter $\lambda^{\mu(s-1)}$. Thus we need no further condition. The longitudinal traceless part is removed by gauge transformation, since $\lambda^{\mu(s-1)}$ is traceful. Therefore, we have no positivity-violating component, though it is unclear whether the lower spin fields as the trace parts can be eliminated.

\section{Equations of motion for higher spin fields}\label{sec:eom}
In the previous section, we have discussed the higher spin symmetry in the kinematical aspect. In other words, what we have shown is that the unitary transformation of the matrices, when translated into terms of derivative operators, includes gauge transformations, and that they remove components of fields except the transverse ones of totally symmetric part.

However, the transformation law for the totally symmetric field is somewhat different from the Fronsdal theory, due to absence of traceless conditions both for the field and for gauge parameter. Therefore there emerges one question: in what form the equations of motion are. Even in the free part, we do not expect it to be the Fronsdal operator. Apparently it conflicts with the existence of higher spin symmetry. In this section we explicitly write down the equations of motions for the field and discuss their structure.

In this paper we truncate the interaction part. It is still worth analyzing since the ordinary higher spin field theory is established rigorously as free field theory.

We shall expand the equations of motion for matrices by substituting Eq.(\ref{eq:gaugefixedA}):
\aln{
0=&\{p_b+A_b,\{p^b+A^b,p^a+A^a\}\}\nonumber\\
\sim&\{p_b,\{p^b,A^a\}\}\nonumber\\
=&\left[\partial^{c(s-2)}\left(\Box h^{a\mu d(s-2)}-2\partial^b\partial^{(a}h_{b}^{~~\mu)d(s-2)}+\partial^a\partial^\mu \bar{h}^{d(s-2)}\right)\right]t_{cd}^{s-2}~p_\mu\nonumber\\
&+\left[\partial^{c(s-1)}\left(\partial^b\partial^ah_b^{~~d(s-1)}-\Box h^{a d(s-1)}\right)\right]t_{cd}^{s-1},
}
where $\bar{h}^{d(s-2)}=h_{b}^{~~bs(s-2)}$. It is remarkable that the coefficients of $t_{cd}^{n}p_{\mu(s-1-n)}$ $(0\leq n\leq s-3)$ vanish, leaving the two equations (neglecting the interaction): 
\aln{
&\partial^{c(s-2)}\left(\Box h^{a\mu d(s-2)}-2\partial^b\partial^{(a}h_{b}^{~~\mu)d(s-2)}+\partial^a\partial^\mu \bar{h}^{d(s-2)}\right)=0,\label{eq:eomh}\\
&\partial^{c(s-1)}\left(\partial^b\partial^ah_b^{~~d(s-1)}-\Box h^{a d(s-1)}\right)=0.\label{eq:eomhsub}
} 
Here we emphasize that the indices of $c$'s and $d$'s are symmetrized respectively, while the two types of indices are antisymmetrized. Note that Eq.(\ref{eq:eomhsub}) is obtained from Eq.(\ref{eq:eomh}) by taking a derivative $\partial^\nu$ and antisymmetrizing $\mu$ and $\nu$. Therefore, we have derived a single equation of motion for the higher spion field.

Eq.(\ref{eq:eomh}) is different from the Fronsdal equation, or equivalently, from the vanishing condition of the Fronsdal operator:
\aln{
\Box h^{a\mu d(s-2)}-2\partial^b\partial^{(a}h_{b}^{~~\mu d(s-2))}+\partial^{(a}\partial^\mu \bar{h}^{d(s-2))}=0.
}
Rather, Eq.(\ref{eq:eomh}) can be understood as a vanishing condition of a kind of curvature. the equation can be written as
\aln{
\eta_{cc}R^{c(s),d(s)}=0, ~~R^{c(s),d(s)}=\partial^{c(s)}h^{d(s)}. 
\label{eq:generalizedcurvature}
}
In the viewpoint of symmetry, $R^{c(s),d(s)}$ corresponds to the Young tableaux of two rows, both of which consist of $s$ boxes. This quantity is the generalization of the (linearized) Riemann curvature, that was discussed in \cite{deWit:1979sib}. It is the only gauge-invariant quantity without the traceless conditions. Thus the appearance of the generalized curvature in the equations of motion is consistent, since we have higher spin symmetry without traceless conditions. Moreover, in $s=2$ case, the above equation is nothing but the Rich-flat condition obtained in Eq.(\ref{eq:eomoriginal}). This coincidence is reasonable because we need neither double-tracelessness for the field, nor the tracelessness for the gauge parameter. in the higher spin case, we conclude that the higher spin field is not the Fronsdal field, but the generalized curvature field.

On the other hand, once we take the interaction into account, the analysis will get far complicated. In the free part of the equations of motion, we obtained the vanishing condition of a derivative operator of degree-$(s-1)$. However, in the presence of the interaction terms coming from products of the second term in Eq.(\ref{eq:gaugefixedA}), the degree of the derivative operator increases, to $2s-3$ at most. In that case, we have many independent equations since all the coefficients of a degree-$(2s-3)$ derivative operator must vanish. Moreover, as long as we consider a single field of spin $s$ only, most of those equations should be regarded as some constraints on the interaction terms. One way to avoid it is to introduce new fields and to make each equation contain free kinetic terms for the field. It is likely that a true consistent description is obtained only when we take into account fields of all spin at the same time. it is equivalent to considering the most general derivative operators of infinite degree, without any truncation.

\section{Summary and discussion}\label{sec:summary}
In this paper, we have studied whether the IIB matrix model contains higher spin fields in its DoF. In particular, we have analyzed the matrix model with the operator interpretation, and have investigated the gauge transformations which emerges from the $U(N)$ transformation of the matrices. Although the class of fields discussed in this paper has been limited, we have found that there is higher spin gauge transformations for them. In order to close the gauge symmetries, we have had to introduce many additional DoF. We can, however, remove the parts of unnecessary and positivity-violating components by suitable gauge transformations. As for the rest parts, we can rewrite them in terms of the totally symmetric part of the original field, $h^{a\mu(s-1)}$. This field is the sole independent DoF of spin-$s$. Furthermore, there is another gauge transformation that removes its longitudinal components. As a result, we have the appropriate spin-$s$ field with transverse components. We also have shown that the field is not Fronsdal field, but the generalized curvature.

There are many aspects that remain to be analyzed. Our present study has been rather qualitative and more concrete analysis is needed. First, the structure of interaction terms has to be analyzed. As we have stated in the last of the previous section, the interaction terms generate numerous term as a derivative operator in the equations of motion, all of which have to vanish at the same time. Although it is desirable to include fields of arbitrary spin and their interactions in the formulation (otherwise there emerge many constraints in the interaction, and they probably leads to inconsistency). Such formulation is too complicated to study by directly expanding matrices as derivative operators. A new method to investigate needs to be established. Related to this issue, it is remarkable that the higher spin gauge transformation in the matrix model includes both homogeneous and inhomogeneous terms. Our study has focused on the inhomogeneous term only, since we have examined whether there are sufficient gauge parameters to eliminate unwanted components. The exact gauge symmetries are far complicated, and it enables the model to include interaction terms. The relationship to the various no-go theorems that prohibit the existence of interacting higher-spin particles needs to be analyzed as well. Further investigation is required. 
The analysis of higher spin symmetries for a general class of fields is another open question. As reviewed in Section \ref{sec:review}, the essential part of the operator interpretation is actually the introduction of the principal bundle. Although we have dealt with the zero modes in the fiber direction, $a^{a|\mu(s-1)}(x,g\!\!\!\!\times)$, the study on symmetries of general fields are a future work. The stability, which seems to be put in danger by higher derivative term in the equations of motions (\ref{eq:generalizedcurvature}), is also an open question.

\section*{Acknowledgement} 
This work is supported by Grant-in-Aid for JSPS Research Fellow Number 17J02185.



\begin{thebibliography}{40}
\bibitem{Weinberg:1980gg} 
  S.~Weinberg,
  General relativity, 790 (1979)



\bibitem{Ishibashi:1996xs} 
  N.~Ishibashi, H.~Kawai, Y.~Kitazawa and A.~Tsuchiya,
  ``A Large N reduced model as superstring,''
  Nucl.\ Phys.\ B {\bf 498}, 467 (1997)
  [hep-th/9612115].


\bibitem{Aoki:1998bq} 
  H.~Aoki, S.~Iso, H.~Kawai, Y.~Kitazawa, A.~Tsuchiya and T.~Tada,
  ``IIB matrix model,''
  Prog.\ Theor.\ Phys.\ Suppl.\  {\bf 134}, 47 (1999)
  [hep-th/9908038].


\bibitem{Fukuma:1997en} 
  M.~Fukuma, H.~Kawai, Y.~Kitazawa and A.~Tsuchiya,
  ``String field theory from IIB matrix model,''
  Nucl.\ Phys.\ B {\bf 510}, 158 (1998)
  [hep-th/9705128].



\bibitem{Aoki:1998vn} 
  H.~Aoki, S.~Iso, H.~Kawai, Y.~Kitazawa and T.~Tada,
  Prog.\ Theor.\ Phys.\  {\bf 99}, 713 (1998)
  [hep-th/9802085].



\bibitem{Iso:1999xs} 
  S.~Iso and H.~Kawai,
  ``Space-time and matter in IIB matrix model: Gauge symmetry and diffeomorphism,''
  Int.\ J.\ Mod.\ Phys.\ A {\bf 15}, 651 (2000)
  [hep-th/9903217].




\bibitem{Kim:2011cr} 
  S.~W.~Kim, J.~Nishimura and A.~Tsuchiya,
  ``Expanding (3+1)-dimensional universe from a Lorentzian matrix model for superstring theory in (9+1)-dimensions,''
  Phys.\ Rev.\ Lett.\  {\bf 108}, 011601 (2012)
  [arXiv:1108.1540 [hep-th]].


\bibitem{Kim:2011ts} 
  S.~W.~Kim, J.~Nishimura and A.~Tsuchiya,
  Phys.\ Rev.\ D {\bf 86}, 027901 (2012)
  [arXiv:1110.4803 [hep-th]].


\bibitem{Kim:2012mw} 
  S.~W.~Kim, J.~Nishimura and A.~Tsuchiya,
  JHEP {\bf 1210}, 147 (2012)
  [arXiv:1208.0711 [hep-th]].




\bibitem{Ito:2015mem} 
  Y.~Ito, J.~Nishimura and A.~Tsuchiya,
  PoS LATTICE {\bf 2015}, 243 (2016)
  [arXiv:1512.01923 [hep-lat]].



\bibitem{Ito:2017rcr} 
  Y.~Ito, J.~Nishimura and A.~Tsuchiya,
  JHEP {\bf 1703}, 143 (2017)
  [arXiv:1701.07783 [hep-th]].



\bibitem{Nishimura:2013moa} 
  J.~Nishimura and A.~Tsuchiya,
  JHEP {\bf 1312}, 002 (2013)
  [arXiv:1305.5547 [hep-th]].



\bibitem{Aoki:2014cya} 
  H.~Aoki, J.~Nishimura and A.~Tsuchiya,
  JHEP {\bf 1405}, 131 (2014)
  [arXiv:1401.7848 [hep-th]].



\bibitem{Steinacker:2007dq} 
  H.~Steinacker,
  JHEP {\bf 0712}, 049 (2007)
  [arXiv:0708.2426 [hep-th]].



\bibitem{Grosse:2008xr} 
  H.~Grosse, H.~Steinacker and M.~Wohlgenannt,
  JHEP {\bf 0804}, 023 (2008)
  [arXiv:0802.0973 [hep-th]].




\bibitem{Steinacker:2008ri} 
  H.~Steinacker,
  Nucl.\ Phys.\ B {\bf 810}, 1 (2009)
  [arXiv:0806.2032 [hep-th]].




\bibitem{Kawai:2016wfh} 
  H.~Kawai, K.~Kawana and K.~Sakai,
  PTEP {\bf 2017}, no. 4, 043B06 (2017)
  [arXiv:1610.09844 [hep-th]].




\bibitem{Sperling:2018xrm} 
  M.~Sperling and H.~C.~Steinacker,
  Nucl.\ Phys.\ B {\bf 941}, 680 (2019)
  [arXiv:1806.05907 [hep-th]].



\bibitem{Sperling:2019xar} 
  M.~Sperling and H.~C.~Steinacker,
  arXiv:1901.03522 [hep-th].




\bibitem{Hanada:2005vr} 
  M.~Hanada, H.~Kawai and Y.~Kimura,
  Prog.\ Theor.\ Phys.\  {\bf 114}, 1295 (2006)
  [hep-th/0508211].

  
\bibitem{Kawai:2007zz} 
  H.~Kawai,
  Prog.\ Theor.\ Phys.\ Suppl.\  {\bf 171}, 99 (2007).


\bibitem{Sakai:2017dxi} 
  K.~Sakai,
  Nucl.\ Phys.\ B {\bf 925}, 195 (2017)
  [arXiv:1709.02588 [hep-th]].



\bibitem{Asano:2012mn} 
  Y.~Asano, H.~Kawai and A.~Tsuchiya,
  Int.\ J.\ Mod.\ Phys.\ A {\bf 27}, 1250089 (2012)
  [arXiv:1205.1468 [hep-th]].


\bibitem{Kawai:2013wwa} 
  H.~Kawai,
  Int.\ J.\ Mod.\ Phys.\ A {\bf 28}, 1340001 (2013).

 


\bibitem{Hamada:2015dja} 
  Y.~Hamada, H.~Kawai and K.~Kawana,
  PTEP {\bf 2015}, no. 12, 123B03 (2015)
  [arXiv:1509.05955 [hep-th]].



\bibitem{Kawana:2016tkw} 
  K.~Kawana,
  arXiv:1609.00513 [hep-th].
 




\bibitem{Fronsdal:1978} 
  C. Fronsdal,
  Phys. Rev. D {\bf 18}, 3624 (1978)



\bibitem{Vasiliev:1980as} 
  M.~A.~Vasiliev,
  Yad.\ Fiz.\  {\bf 32}, 855 (1980)
  [Sov.\ J.\ Nucl.\ Phys.\  {\bf 32}, 439 (1980)].




\bibitem{Didenko:2014dwa} 
  V.~E.~Didenko and E.~D.~Skvortsov,
  arXiv:1401.2975 [hep-th].



\bibitem{Engquist:2007yk} 
  J.~Engquist and O.~Hohm,
  JHEP {\bf 0804}, 101 (2008)
  [arXiv:0708.1391 [hep-th]].



\bibitem{deWit:1979sib} 
  B.~de Wit and D.~Z.~Freedman,
  Phys.\ Rev.\ D {\bf 21}, 358 (1980).
























  





\end{thebibliography}
\end{document}